\documentclass[12pt]{article}
\usepackage{amsfonts}
\usepackage{amssymb}
\usepackage{amsmath}
\usepackage{amsthm}
\usepackage{amsfonts}
\usepackage{multicol}
\usepackage{amscd}
\usepackage{textcomp}
\usepackage{multirow}
\usepackage[english, activeacute]{babel}

\def\be{\begin{equation}}
\def\ee{\end{equation}}
\def\bea{\begin{eqnarray}}
\def\eea{\end{eqnarray}}

\newcommand{\sect}[1]{\setcounter{equation}{0}\section{#1}}
\newcommand{\subsect}[1]{\subsection{#1}}
\newcommand{\subsubsect}[1]{\subsubsection{#1}}

\def\1{\'{\i}}

\def\>#1{{\mathbf#1}}

\def\rr{{\cal R}}
\def\sss{{\cal S}}

\def\la{\lambda}

\def\ene{N}
\def\eme{N}

\def\kk{K}

\begin{document}

\thispagestyle{empty}

\hfill \

\ 
\vspace{0.5cm}

\begin{center}

{\Large{\sc{N-dimensional integrability from two-photon\\[4pt]
 coalgebra symmetry}} }

\end{center}

\medskip

\begin{center} \'Angel Ballesteros, Alfonso Blasco and Francisco J.
Herranz 
\end{center}

\begin{center} {\it {Departamento de F\1sica,  Universidad de Burgos, 
09001 Burgos, Spain}}
\end{center}

  \medskip

\begin{abstract} 
\noindent
A wide class of Hamiltonian systems with $N$ degrees of freedom and endowed with, at least, $(N-2)$ functionally independent integrals of motion in involution is constructed by making use of the two-photon Lie--Poisson coalgebra $(h_6,\Delta)$. The set of $(N-2)$ constants of the motion is shown to be a universal one for all these Hamiltonians, irrespectively of the dependence of the latter on several arbitrary functions and $N$ free parameters. Within this large class of quasi-integrable $N$-dimensional Hamiltonians,  new families of completely integrable  systems are identified by finding explicitly a new independent integral ${\cal I}$ through the analysis of the sub-coalgebra structure of $h_6$. In particular, new  completely integrable $N$-dimensional Hamiltonians describing natural systems, geodesic flows and static electromagnetic Hamiltonians are presented.

 \end{abstract}

\bigskip\bigskip\bigskip\bigskip



\vfill
\newpage


\noindent

\sect{Introduction}

Due to its physical and mathematical relevance, the construction of completely integrable Hamiltonian systems focuses an intense research activity that makes use of many different approaches and techniques (see, for instance,~\cite{Perelomov,Audin, Bolsinov, Babelon}). However, the number of known integrable systems (in the Liouville sense~\cite{AKN97}) that can be generalized for an arbitrary number $N$ of degrees of freedom is relatively scarce. In most of the cases, this possibility is based on some underlying symmetry that allows for the appropriate propagation of the integrability properties to arbitrary dimension (see the systems described in~\cite{Garnier}--\cite{Verrier} and references therein).

The aim of this paper is the construction  
of many new families of classical integrable Hamiltonian systems with $N$ degrees of freedom, depending on several arbitrary functions and $N$ free parameters, and whose integrals of the motion will be also explicitly given. These results will be obtained by applying the coalgebra symmetry method (see~\cite{BR, CRMAngel, BHnonlin}) to the Poisson version of the so-called two-photon/Schr\"odinger algebra $h_6$~\cite{Gil, Brif, poissonh6}. As we shall show in the sequel, the $h_6$-coalgebra symmetry turns out to be extremely powerful, since the $N$-particle realization provided by the coalgebra structure encompasses different $N$-body symmetries that underly several useful and explicit integrability properties. Moreover, one can easily realize that the approach here presented is also applicable in order to ensure the integrability of the quantum mechanical analogues of all the Hamiltonians contained in this paper.

Firstly, let us recall that the $h_6$ Lie--Poisson algebra is
spanned  by the  six abstract generators
$\{\kk,A_+,A_-,B_+,B_-,M\}$, whose one-particle symplectic realization is given by
\be 
\begin{array}{lll}
A_+=\la_1 \, p_1&\quad
A_-=\la_1  \,q_1 &\quad   \displaystyle{
\kk=q_1\,  p_1  -\frac  {\la_1^2}{2} }\cr
\displaystyle{ B_+= {p_1^2}}
&\quad B_-= 
q_1^2   &\quad   M=\la_1^2 
\end{array}
\label{ad}
\ee 
where $\la_1$ is a non-vanishing constant that labels the previous symplectic realization
and where we have considered the usual Poisson bracket $\{q_1,p_1\}=1$.  The abstract Poisson brackets defining $h_6$ read
\be 
\begin{array}{lll}
\{\kk,A_+\}=A_+&\qquad \{\kk,A_-\}=-A_- &\qquad \{A_-,A_+\}=M\cr
\{\kk,B_+\}=2B_+&\qquad \{\kk,B_-\}=-2 B_- &\qquad \{B_-,B_+\}=4\kk+2M\cr
\{A_+,B_-\}=-2 A_-  & \qquad \{A_+,B_+\}=0& \qquad
\{M,\,\cdot\,\}=0\cr
\{A_-,B_+\}=2A_+&\qquad \{A_-,B_-\}= 0.& 
\end{array}
\label{aa}
\ee
A direct inspection of this algebra makes its rich subalgebra structure evident (for instance, the $gl(2)$ subalgebra $\{\kk,B_+,B_-,M\}$, the oscillator one $h_4\equiv\{\kk,A_+,A_-,M\}$ and the Heisenberg one $h_3\equiv\{A_+,A_-,M\}$ can be easily identified). This is one of the main features of  $h_6$, since this algebra generalizes many lower dimensional Lie symmetries in a transparent way, a fact that will be relevant in order to find new and more general integrability structures.

As a consequence of (\ref{ad}), any quadratic Hamiltonian
with one degree of freedom can  always be written as a linear combination
of the $h_6$ generators in the above representation. This is indeed the origin of the quantum mechanical relevance of the $h_6$ Lie algebra as a dynamical symmetry, since the quantum
counterpart of such a quadratic Hamiltonian can be interpreted as a
single-mode radiation field Hamiltonian including the number operator $\hat\kk$,
creation and annihilation operators ${\hat A}_+$ and ${\hat A}_-$ and two-photon creation
and annihilation operators
${\hat B}_+$ and ${\hat B}_-$ (the generator ${\hat M}$ would be a central one). In this context, different applications of the $h_6$ symmetry can be found in~\cite{Brif}, and it is also interesting to recall that $h_6$ is isomorphic to
the $(1+1)$-dimensional centrally extended Schr\"odinger Lie algebra~\cite{twop}. In fact, this
isomorphism  provides the two Casimir functions for $h_6$, which are the central
generator $M$ and the fourth-order Casimir~\cite{Patera}
\be
{\cal C}= (MB_+ -A_+^2)(MB_- - A_-^2)-
(M \kk -   A_- A_+ +M^2/2)^2 
\label{aac}
\ee
that will play a relevant role hereafter. The function ${\cal C}$ can be reduced to a third-order invariant by extracting $M$ as a
common factor:
\be
{\cal C}_{h_6}={\cal C}/{M}  
 =MB_+B_- - B_+ A_-^2  - B_- A_+^2  - M(\kk+M/2)^2 + 2
A_-A_+(\kk+M/2)  .
\label{ac}
\ee 
Note that if we substitute the symplectic realization (\ref{ad}) within the two Casimir functions, the former is characterized by the values $M=\la_1^2$ and ${\cal
C}_{h_6}=0$.  

Now, if we endow the $h_6$ Poisson algebra with a coalgebra structure, {\em i.e.} with a two-body primitive coproduct (which is  a Poisson algebra
homomorphism
$\Delta: h_6\rightarrow h_6\otimes h_6$ \cite{BR,BHnonlin}) given by
\be
\Delta(X)= X \otimes 1 +1\otimes X  
\qquad X\in \{\kk,A_+,A_-,B_+,B_-,M\} ,
\label{ab}
\ee
then the one-particle dynamical symmetry given by (\ref{ad}) can be extended recursively. This gives rise to an infinite family of $N$-particle Hamiltonians defined as {\em any} smooth function ${\cal H}$ of the generators of $h_6$ 
$$
{\cal H}={\cal H}(\kk ,B_+ ,B_- ,A_+ ,A_- ,M )
\label{hintro}
$$
and provided that ${\cal H}$ is realized in the $N$-particle symplectic realization coming from the coproduct structure. 
More importantly, the $m$-particle realizations ($m=3,\dots,N$) of the coproducts of the Casimir function ${\cal C}_{h_6}$ will provide a set of $(N-2)$ independent integrals of the motion in involution with the whole family of Hamiltonians ${\cal H}$ (as we shall explain later, the $m=2$ realization gives a vanishing constant of the motion). Moreover, since  there is only {\em one} integral of motion left in order to get the {\em complete integrability} of ${\cal H}$, we shall name all these systems as {\em
quasi-integrable} Hamiltonians.  This construction will be described in detail in section 2, and some preliminary results can be found in~\cite{BHnonlin}. Moreover, having in mind further applications, particular choices for ${\cal H}$ leading to $N$-dimensional  ($N$D) natural systems, geodesic flows and static electromagnetic Hamiltonians will be explicitly identified. 
 
In section 3 we analyse two different possibilities in order to ensure the complete integrability of ${\cal H}$ by finding an extra integral ${\cal I}$, that will be defined as a function of the generators of $h_6$, since this guarantees its existence for any number of degrees of freedom. In the first case,
the Hamiltonian can be defined on a {\em
subalgebra} of $h_6$ and then the $N$-particle Casimir of such subalgebra can be directly identified with  the additional constant $\cal I$. This possibility is presented in section 4, where the rich subalgebra structure for $h_6$ is fully described. 

A second possibility is analysed in section 5, where we introduce five new families of $N$D integrable systems that have one of the five non-central generators of $h_6$  as the remaining integral of  motion $\cal I$. This construction turns out to be significantly powerful, since all these systems depend on several arbitrary functions and provide a large number of new instances of interesting Hamiltonians, both from the mathematical and the physical viewpoints.
Finally, some comments and further research lines are presented in a concluding section.


\sect{An infinite family of quasi-integrable systems}

The integrability properties of the generic $N$-particle Hamiltonians with $h_6$ coalgebra symmetry relies on the following result.

\medskip

\noindent
{\bf Theorem 1.}
{\it Let $\{\>q,\>p \}=\{(q_1,\dots,q_N),(p_1,\dots,p_N)\}$ be $N$ pairs of
canonical variables. The $N$D Hamiltonian
\begin{equation}
{\cal H}={\cal H}(\kk ,B_+ ,B_- ,A_+ ,A_- ,M )
\label{hgen}
\end{equation}
defined as any smooth function ${\cal H}: \mathbb R^6 \rightarrow \mathbb R$ and
\bea
&& A_+ =\sum_{i=1}^N \la_i p_i \qquad
 A_- =
\sum_{i=1}^N \la_i  q_i\qquad   \kk = \sum_{i=1}^N
\bigg(q_i p_i -\frac {\la_i^2}2\bigg)  \cr 
&& B_+ =\sum_{i=1}^N 
{p_i^2} \qquad\!\quad
 B_- = 
\sum_{i=1}^N  q_i^2\qquad\,\,
 M =\sum_{i=1}^N
\la_i^2   ,
\label{sympn}
\eea 
where $\la_i$ are $N$ arbitrary parameters, is 
quasi-integrable. The  $(N-2)$ functionally
independent  integrals of the motion for ${\cal H}$ are
\be
C^{(m)}=
{\sum_{1\leq i<j<k}^{m}} 
\bigl(
\la_i (p_j q_k - p_k q_j ) +
\la_j (p_k q_i - p_i q_k ) +
\la_k (p_i q_j - p_j q_i ) 
\bigr)^2
\label{c3m}
\ee
where $m=3,\dots, N$. These integrals are in involution and can be called `universal' in the sense that they do not depend on the specific choice of the function ${\cal H}$.}
\medskip

 \noindent
{\bf Proof.} The keystone to prove this result comes from the fact that (\ref{ad}) is a one-particle symplectic realization for the Poisson coalgebra $(h_6,\Delta)$, labelled by the $\la_1$ parameter. 
Moreover, it can be easily checked that (\ref{sympn}) is just the  
$N$-particle symplectic realization of $(h_6,\Delta)$ that is obtained through the $N$-sites generalization of the coproduct (\ref{ab}):
\bea
&&\!\!\!\!\!\!\!\!\Delta^{(N)}(X)=X\otimes 1\otimes 1\otimes
\dots^{N-1)}\otimes 1 \cr
&&\qquad\qquad\quad + 1\otimes X\otimes 1\otimes\dots^{N-2)}\otimes
 1 +\dots \cr
&&\qquad\qquad\qquad\quad + 1\otimes
1\otimes\dots^{N-1)}\otimes 1\otimes X .
\label{ca}
\eea
This means that the $N$-particle generators
(\ref{sympn}) fulfil the  commutation rules (\ref{aa}) with respect to the canonical Poisson
bracket 
$$
\{f,g\}=\sum_{i=1}^N\left(\frac{\partial f}{\partial q_i}
\frac{\partial g}{\partial p_i}
-\frac{\partial g}{\partial q_i} 
\frac{\partial f}{\partial p_i}\right).  
\label{bg}
$$

Moreover, such coalgebra symmetry expressed through the symplectic realization implies that each of the $N$-particle generators (\ref{sympn}) Poisson-commute with the
$(N-1)$ functions  $C^{(m)}$ given by the $m$-th coproducts of the Casimir (\ref{ac}) with $m=2,3,\dots,N$ (see~\cite{BR,BHnonlin} for details). However, in the case of $h_6$ the $C^{(2)}$ function vanishes (the two-body coproduct of the Casimir is zero~\cite{BHnonlin}), and we are left with the set of $(N-2)$ integrals (\ref{c3m}) that, also by construction, are functionally independent and Poisson-commuting. Therefore, any function ${\cal H}$ (\ref{hgen}) of the $N$-particle symplectic realization of the $h_6$ generators will be in involution with the set of integrals $C^{(m)}$, which completes the proof.

\medskip

This quite general  result deserves the following remarks and comments:
\begin{itemize}

\item We can properly say that ${\cal H}$ is a {\em quasi-integrable} Hamiltonian, since for {\em any dimension} $N$ and {\em any choice of ${\cal H}$} there is only one integral left in order to get its complete integrability. Obviously, some specific choices for ${\cal H}$ will lead to {\em completely integrable} Hamiltonians for which an additional integral does exist for any dimension $N$. The aim of this paper is just to find solutions to this problem.

\item Within the coalgebra approach it is well-known that, in general, two different sets of integrals of the motion coming from `left' and `right'
$m$-th coproducts of the Casimir can be obtained  (see \cite{CRMAngel} for
details). Indeed, this is also the case for the $(h_6,\Delta)$ coalgebra, where by making use of  the `right'
$m$-th coproducts, the following alternative set of $(N-2)$ integrals in involution $C_{(m)}$ is obtained:
\be
C_{(m)}=
{\sum_{ N-m+1\le i<j<k,}^{N}} 
\bigl(
\la_i (p_j q_k - p_k q_j ) +
\la_j (p_k q_i - p_i q_k ) +
\la_k (p_i q_j - p_j q_i ) 
\bigr)^2 .
\label{c3mr}
\ee
This means that if we label the $N$ sites on
$h_6\otimes h_6\otimes  \dots^{N)}\otimes h_6$ by
$1\otimes 2\otimes\dots\otimes N$, the `left' Casimir  
 ${ C}^{(m)}$  is defined on the sites $1\otimes 2\otimes \dots \otimes m$, while
the `right' one ${ C}_{(m)}$  is defined on $(N-m+1)\otimes \dots \otimes N-1\otimes N$.
Moreover, it is straightforward to prove that the   $(2N-4)$ functions $\{ { C}^{(3)},{ C}^{(4)},\dots ,{ C}^{(N)}\equiv { C}_{(N)}, {
C}_{(N-1)},\dots { C}_{(3)},{\cal H}
\}$ are functionally independent (assuming that ${\cal H}$ is not a function of ${\cal C}$ only) and the coalgebra symmetry ensures that each of the subsets
$\{{ C}^{(3)},\dots,{ C}^{(N)},{\cal H} \}$ or $\{{ C}_{(3)},\dots,{
C}_{(N)},{\cal H} \}$ is formed by $(N-1)$  functions in involution~\cite{BR,CRMAngel}. 

\item As a consequence, in case that an additional integral ${\cal I}$ is found for a given ${\cal H}$, this Hamiltonian will be not only integrable, but also {\em superintegrable} provided that the $(N-3)$ `right' constants ${  C}_{(m)}$ (with $m=3,\dots, N-1$) commute with ${\cal I}$ (this fact will be certainly ensured if ${\cal I}$ is a function of the $N$-particle symplectic realization of the $h_6$ coalgebra). 

\item The role of the $\la_i$ parameters is essential in this approach, since they provide an $N$-parameter freedom for the Hamiltonian. From a co-algebraic viewpoint these $\la_i$ parameters can be neatly interpreted: each of them fixes the one-particle symplectic realization that we are using on the $i$-th site of the underlying $h_6$ symmetry lattice $h_6\otimes h_6\otimes  \dots^{N)}\otimes h_6$.

\item We stress that  the integrals   (\ref{c3m}) can be interpreted as sums of the squares of a linear combination (through the $\la_i$ parameters) of   `Euclidean angular momentum' components $J_{ij}$. In particular the $N(N-1)/2$ functions $J_{ij}=q_i p_j -q_j p_i$ with $i<j$ and $i,j=1,\dots ,N$ span an ${so}(N)$ Lie--Poisson algebra so that ${C}^{(m)}$ can be read as
$$
C^{(m)}=
{\sum_{1\le i<j<k}^{m}}  
\bigl(
\la_i J_{jk} +
\la_j J_{ki}+
\la_k J_{ij}
\bigr)^2 
$$
where $J_{ki}=-J_{ik}$. Hence each term in $C^{(m)}$ is the square of an element of the Lie--Poisson algebra ${so}(3)=\{ J_{ij}, J_{ik}, J_{jk}\}$ (and the same happens in ${C}_{(m)}$). From this perspective, the $h_6$-coalgebra symmetry  can be interpreted as a `generalization'  of the spherical symmetry, which will be fully recovered when the Hamiltonian is defined on the $gl(2)$ Poisson algebra.

\item Although the central generator $M$ is also a Casimir for the two-photon coalgebra, its $N$-th coproduct  gives  rise to $N$ trivial integrals of the motion
\be 
M^{(m)}=\sum_{i=1}^m
\la_i^2  
\qquad m=1,\dots,N ,
\label{cd}
\ee
that do not provide any dynamical information. In this sense, $M$ can be considered either as a generator on its own right or as a constant that depends on both the dimension $N$ and the chosen symplectic realizations through the $\la_i$ parameters. 

\item Finally, is worth mentioning that the role of the $h_6$ algebra in the integrability properties of certain 3D Hamiltonian systems was already pointed out from a different viewpoint in~\cite{Rauchh6}. Nevertheless, the introduction of a coalgebra structure in $h_6$ turns out to be essential in order to fully exploit its integrability information and to generalize them to arbitrary dimensions.
  
\end{itemize}


\subsect{Some relevant quasi-integrable Hamiltonians}

Among the bunch of $N$D quasi-integrable systems that are provided by theorem 1, the following particular classes are physically outstanding.

\subsubsect{Natural systems}

The Hamiltonian
\be
{\cal
H}=\frac 12 {B_+} +  {\cal F}\left( {A_-},B_-\right)
\label{eucl}
\ee
where ${\cal F}$ is a   function playing the role of a potential, gives rise to the following quasi-integrable system on the $N$D Euclidean space $\mathbb E^N$:
\be
{\cal H}=\sum_{i=1}^\ene\frac {p_i^2}{2}+{\cal
F}\biggl(\sum_{i=1}^\ene\la_i q_i, 
\sum_{i=1}^\ene q_i^2\biggr).  
\label{eucla}
\ee
Notice that central potentials, with spherical symmetry, directly arise whenever $ {\cal F}$ does not depend on $A_-$ (the Hamiltonian is then defined on the $gl(2)$ subalgebra). Thus in the case with generic $ {\cal F}\left( {A_-},B_-\right)$, the  spherical symmetry is broken and  its associated (super)integrability is, in principle, reduced to quasi-integrability.

\subsubsect{Electromagnetic Hamiltonians} 

The most general $N$D quasi-integrable Hamiltonian including linear terms in the momenta is given by
\be
{\cal
H}=\frac 12 {B_+}  + \kk\,{\cal F}\left( {A_-},B_-\right) + {A_+}\,{\cal G}\left(
{A_-},B_-\right) + \rr\left( {A_-},B_-\right)
\ee
where   ${\cal F}$, ${\cal G}$ and $\rr$ are arbitrary smooth functions. In terms of canonical variables, it reads
\bea
&& {\cal
H} =\sum_{i=1}^\ene\frac {p_i^2}{2}
+\left(\sum_{i=1}^\ene{\left(q_i p_i -\frac {\la_i^2}2\right)}\right) {\cal
F}\biggl(\sum_{i=1}^\ene\la_i q_i, 
\sum_{i=1}^\ene q_i^2\biggr)  \cr
&&
\qquad\qquad+\left( \sum_{i=1}^\ene \la_i p_i
\right) {\cal
G}\biggl(\sum_{i=1}^\ene\la_i q_i, 
\sum_{i=1}^\ene q_i^2\biggr)  +
\rr\biggl(\sum_{i=1}^\ene\la_i q_i, 
\sum_{i=1}^\ene q_i^2\biggr) .
\label{electro}
\eea
In 3D, this Hamiltonian describes the motion of a particle on $\mathbb E^3$ under the action of a static electromagnetic field which is determined by the functions ${\cal F}$, ${\cal G}$ and $\rr$. 
Namely, if we compare (\ref{electro}) with the 3D electromagnetic Hamiltonian 
\be
\mathcal{H}_{em}=\dfrac{1}{2}\left(\vec p-e\vec A\right)^{2}+e\, \psi
\ee
where $e$ is the electric charge, $\vec A$ is the vector potential and $\psi$ is the scalar one, 
we get
\bea
&&
\!\!\!\!\!\!\!\!\!\!\!\!\!\!
A_{i}=-\dfrac{q_{i}}{e}\mathcal{F}\left(A_{-},B_{-}\right)-\dfrac{\la_{i}}{e}\mathcal{G}\left(A_{-},B_{-}\right)\qquad i=1,2, 3
\\
&& \!\!\!\!\!\!\!\!\!\!\!\!\!\!
\psi=\dfrac{1}{e}\mathcal{R}\left(A_{-},B_{-}\right)-\dfrac{1}{2e}M\,\mathcal{F}(A_{-},B_{-})
\nonumber\\
&& 
-\dfrac{1}{2e}\left[
B_{-}\mathcal{F}\left(A_{-},B_{-}\right)^{2}+2A_{-} \mathcal{F}(A_{-},B_{-})\mathcal{G}(A_{-},B_{-})+M\mathcal{G}(A_{-},B_{-})^{2}
\right] .
\eea
Note the relevant role that the $\la_i$ parameters play in the definition of the electromagnetic field.
Recall also that $N$D superintegrable  electromagnetic systems have been recently obtained in~\cite{BH07} by making use of an ${sl}(2,\mathbb R)$-coalgebra symmetry. In fact, as we shall show in section 4, the latter systems are a particular subfamily of (\ref{electro}) since ${sl}(2,\mathbb R)$ is a sub-coalgebra of $h_6$. We also recall that 2D integrable  electromagnetic Hamiltonians had
been previously studied in \cite{Hietarinta,Dorizzi,winter}.

\subsubsect{Geodesic flow Hamiltonians} 

A third family of relevat systems is the one given by $N$D quasi-integrable Hamiltonians of the type 
\be
{\cal H}=\sum_{i,j=1}^N{g^{ij}(q_1,\dots,q_N)\,p_i\,p_j}
\ee
that are obtained by considering 
\begin{eqnarray}
&& {\cal 
H}=B_{+}\mathcal{F}(A_{-},B_{-})+A_{+}^2\mathcal{G}(A_{-},B_{-}) \notag \\
&& \qquad\qquad +\left(\kk+\dfrac{M}{2}\right)^{2} \rr(A_{-},B_{-}) +
 \left(\kk+\dfrac{M}{2}\right)A_{+} \sss(A_{-},B_{-}) 
 \label{free}
\end{eqnarray}
since for any choice of the functions ${\cal F}$, ${\cal G}$, $\rr$ and $\sss$ we obtain a Hamiltonian which is a quadratic homogeneous function in the momenta. Explicitly,
\begin{eqnarray}
&&
{\cal H}=\  \left(\sum_{i=1}^\ene\frac {p_i^2}{2}\right)\, 
\mathcal{F}\left(\sum\limits_{i=1}^{N}\,\la_{i}q_{i},\sum\limits_{i=1}^{N}q_{i}^{2}\right) +\left(\sum\limits_{i=1}^{N}\la_{i}p_{i}\right)^{2}\mathcal{G}\left(\sum\limits_{i=1}^{N}\,\la_{i}q_{i},\sum\limits_{i=1}^{N}q_{i}^{2}\right)  \notag \\
&& \qquad \qquad
+\left(\sum\limits_{i=1}^{N}q_{i}p_{i}\right)^{2} \rr\left(\sum\limits_{i=1}^{N}\,\la_{i}q_{i},\sum\limits_{i=1}^{N}q_{i}^{2}\right)\notag \\
&&\qquad\qquad+\left(\sum\limits_{i=1}^{N}q_{i}p_{i}\right)\left(\sum\limits_{i=1}^{N}\la_{i}p_{i}\right) \sss\left(\sum\limits_{i=1}^{N}\,\la_{i}q_{i},\sum\limits_{i=1}^{N}q_{i}^{2}\right) .
\label{geod}
\end{eqnarray}
We stress that the specific form of the metric $g^{ij}$ is defined by the ${\cal F}$, ${\cal G}$, $\rr$ and $\sss$ functions which, in general, give rise to an $N$D space of {\em nonconstant} curvature.
In any case, the set of constants of   motion (\ref{c3m})  is universal and does not depend on the specific choice of the arbitrary functions in the Hamiltonian. 

On the other hand, we recall that the complete integrability of a free Hamiltonian on a curved space is a rather nontrivial property which is connected with geometric and topological features of the underlying
manifold (see~\cite{Darboux,KH95,Pa99,PLBEnciso}).
Moreover, additional potentials on these $h_6$-coalgebra spaces can be naturally considered   by adding  functions  such as, e.g., ${\cal U}(A_-,B_-)$ to the free Hamiltonian (\ref{free}). In this way the Euclidean natural systems (\ref{eucl}) can be generalized to the curved spaces defined through (\ref{free}) without breaking the quasi-integrability of the geodesic flow Hamiltonian.


\sect{Complete integrability}

At this point the main problem to be faced is the characterization of those Hamiltonians ${\cal H}$ for which 
an additional integral ${\cal I}$ does exist {\em for any dimension $N$}, thus providing their complete $N$D integrability.

In order to ensure the existence of ${\cal I}$ for any dimension $N$, we shall assume that this additional integral is also $h_6$-coalgebra invariant, which means that it can be written as a function
\be
{\cal I}={\cal I}(\kk ,B_+ ,B_- ,A_+ ,A_- ,M )
\label{Iinv}
\ee
where the $h_6$ generators are written in their $N$-particle symplectic realization (\ref{sympn}). In this way, if  ${\cal I}$ is functionally independent with respect to both the $h_6$ Casimir (\ref{ac}) and the Hamiltonian ${\cal H}$, 
the coalgebra symmetry ensures --by construction-- the involutivity of ${\cal I}$ with respect to the $(N-2)$ `left' integrals $C^{(m)}$ $(m=3,\dots,N)$ and its  functional independence with respect to them. And the very same result holds for the 
  $(N-3)$  `right'  integrals $C_{(m)}$ where $m=3,\dots,N-1$ (we recall that $C_{(N)}=C^{(N)}$).  
  
This means that if ${\cal I}$ does exist in the form (\ref{Iinv}), then ${\cal H}$ will be not only a completely integrable system  but also a {\em superintegrable} one, since a total number of $(N-2)+(N-3)+1=(2\,N - 4)$ functionally independent constants of  motion for  ${\cal H}$ has been explicitly found. Nevertheless, even in this superintegrable case, ${\cal H}$ is not a {\em maximally superintegrable} Hamiltonian, since two more independent constants of the motion would be needed to get the maximum possible total number of $(2\,N - 2)$ independent integrals. Again, these two remaining integrals could exist for some very particular choices for ${\cal H}$, but in any case neither their existence nor their explicit form can be derived from the $h_6$ coalgebra symmetry.

The rest of the paper is devoted to show how the search for the additional ${\cal I}$ (\ref{Iinv}) can be guided by taking into account the subalgebra structure of $h_6$. In particular, we shall consider two different situations in which the existence of ${\cal I}$ is guaranteed  by construction:

\medskip

\noindent {\bf A)}  If the Hamiltonian ${\cal H}$ is defined within a subalgebra of $h_6$ that has a non-trivial Casimir invariant, the $N$-particle realization of the Casimir of the subalgebra provides the integral ${\cal I}$. This {\em subalgebra integrability} approach will be analysed in the next section, in which the subalgebra structure for $h_6$ will be fully described.

\medskip

\noindent {\bf B)} Let $X$ be a fixed generator of $h_6$. The $N$-particle symplectic realization of $X$ will Poisson commute with any $N$-particle Hamiltonian ${\cal H}_X$ defined as a function of all the remaining $h_6$ generators commuting with $X$ and of the Casimirs of all the subalgebras containing the given generator $X$. Under such hypotheses, ${\cal H}_X$ is completely integrable since the generator $X$ is just the additional constant of   motion ${\cal I}$. We have five relevant generators $\{\kk ,B_+ ,B_- ,A_+ ,A_- \}$ (the central generator $M$ would give no dynamical information), so this {\em generator integrability} procedure will give rise to five families of completely integrable systems that will be   studied in detail in section 5.

\medskip

Finally, we stress that if a given Hamiltonian does not fit within the two previous approaches, the search for the remaining integral ${\cal I}$ --in case it does exist-- have to be performed by using direct methods. Indeed, some particular solutions can be found, and a particular example will be given in the final section.


\sect{Subalgebra integrability}

The subalgebras of $h_6$ with a non-trivial ({\em i.e.} linear) Casimir  function are summarized in table 1 together with their 1D symplectic realization. They are:

\noindent -- Two `book'  algebras  ${\cal D}_+$ and ${\cal D}_-$   generated by a dilation plus two
translations.

\noindent -- The harmonic oscillator algebra $h_4$.

\noindent -- Two centrally extended $(1+1)$D Galilean
algebras  $\overline{\cal G}_+$ and
$\overline{\cal G}_-$.

\noindent -- A centrally extended 2D Euclidean 
algebra $\overline{\cal E}$ (where $\mu$ and $\nu$ are non-zero real parameters).

\noindent -- The $gl(2)$ algebra. 

More details on these subalgebras and on  their associated Lie--Poisson  structures can be found in~\cite{poissonh6}.
Clearly  the
Heisenberg--Weyl algebra $h_3=\{ A_+,A_-,M\}$ is a subalgebra of $h_4$, and
   $gl(2)$ contains an $sl(2,\mathbb R)$  subalgebra (by mapping  $\kk\to \kk +M/2$), so we
have the following subalgebra embeddings:
\be
h_3\subset h_4\subset h_6\qquad
\overline{\cal G}_\pm\subset h_6\qquad
sl(2,\mathbb R)\subset gl(2)\subset h_6 .
\label{ck}
\ee
Notice also that  $\overline{\cal E}$ is a proper Euclidean subalgebra whenever $\mu$ and $\nu$ have the same sign; on the contrary, $\overline{\cal E}$ is in fact a centrally extended $(1+1)$D Poincar\'e subalgebra. In the sequel we do not distinguish the two  real forms as the resulting expressions for $\overline{\cal E}$ will be  globally parametrized through   $\mu$ and $\nu$.

\begin{table}[ht]
\caption{{Relevant subalgebras of $h_6$.}}
\label{Table1}
 \begin{center}
\noindent
\begin{tabular}{clll}
\hline
\\[-0.3cm]
\multicolumn{1}{l}{Subalgebra}&
\multicolumn{1}{l}{Generators}&
\multicolumn{1}{l}{Symplectic realization }
&\multicolumn{1}{l}{Casimir function}\\[0.1cm]
\hline
\\[-0.3cm]
${\cal D}_+$&$\kk,A_+,B_+$&  $qp-\frac {\la^2}2,
\la p, {p^2}$&$A_+^2/B_+$\\[0.2cm]
${\cal D}_-$ &  $\kk,A_-,B_-$&  $qp-\frac {\la^2}2,   
\la q,  q^2$&$A_-^2/B_-$\\[0.2cm] 
$h_4$&  $\kk,A_-,A_+,M$&  $qp-\frac {\la^2}2,\la
 q,\la p,\la^2$&$M(\kk+\frac 12M)-A_-A_+$ \\[0.2cm] 
$\overline{\cal G}_+$&$B_+,A_-,A_+,M$&  ${p^2},
\la  q,\la p,\la^2$&$MB_+ - A_+^2$\\[0.2cm]
$\overline{\cal G}_-$ &  $B_-,A_-,A_+,M\qquad$&  ${}q^2,
\la  q,\la p,\la^2$&$MB_- - A_-^2$\\[0.2cm] 
$\overline{\cal E}$&  $\mu B_+ + \nu B_-,$&  $\mu 
{p^2}+ \nu q^2,$&$M(\mu B_+ + \nu
B_-)$ \\[0.2cm] 
$\ $&  $A_-,A_+,M$&  $\la  q,\la p,\la^2$&$\quad -\mu A_+^2 - \nu A_-^2$ \\[0.2cm] 
$gl(2)$&  $\kk,B_-,B_+,M$&  $qp-\frac
{\la^2}2,q^2,{p^2},\la^2$&$B_-B_+-(\kk+\frac 12
M)^2$
\\[0.1cm]
\hline
\end{tabular}
\end{center}
\end{table}
 


\begin{table}[ht]
\caption{{$N$D symplectic realization of the   Casimir of each of the sub-coalgebras of $(h_6,\Delta)$ given in table~\ref{Table1}.}}
\label{Table2}
\begin{center}
\noindent
\begin{tabular}{ll}
\hline
\\[-0.3cm]
\multicolumn{1}{l}{Sub-coalgebra}&
\multicolumn{1}{l}{Integrals of   motion}\\[0.1cm]
\hline
\\[-0.3cm]
$({\cal
D}_+,\Delta)$&$\displaystyle{C^{(\eme)}_{{\cal D}_+}=
\biggl( \sum_{{i=1}}^\eme \la_ip_i \biggr)^2/\biggl(
\sum_{{j=1}}^\eme p_j^2 \biggr)}$\\[0.2cm]
$({\cal
D}_-,\Delta)$&$\displaystyle{C^{(\eme)}_{{\cal D}_-}=
\biggl( \sum_{{i=1}}^\eme \la_i q_i \biggr)^2/\biggl(
\sum_{{j=1}}^\eme
 q_j^2 \biggr)}$\\[0.2cm]
$(h_4,\Delta)$&  $\displaystyle{C^{(\eme)}_{h_4}=
\sum_{{1\le i< j}}^{\eme} (\la_j p_i-\la_i p_j)
(\la_jq_i-  \la_iq_j)}$\\[0.2cm] 
$(\overline{\cal
G}_+,\Delta)$&$\displaystyle{C^{(\eme)}_{\overline{\cal G}_+}=
\sum_{{1\le i<j}}^\eme (\la_j p_i -\la_i p_j)^2}
$\\[0.2cm]
$(\overline{\cal G}_-,\Delta)$ & 
$\displaystyle{C^{(\eme)}_{\overline{\cal G}_-}= 
\sum_{{1\le i<j}}^\eme  (\la_j q_i- \la_i q_j)^2 }$\\[0.2cm] 
$(\overline{\cal
E},\Delta)$&$\displaystyle{C^{(\eme)}_{\overline{\cal E}}=
\sum_{{1\le i<j}}^\eme \bigg\{\mu  (\la_jp_i -\la_ip_j)^2 +\nu (\la_j
q_i-
\la_i q_j)^2 \bigg\} }\qquad $\\[0.2cm]
$(gl(2),\Delta) \quad$&  $\displaystyle{
C^{(\eme)}_{gl(2)}=  
\sum_{{1\le i<j}}^\eme (q_j p_i -
q_i p_j)^2 }$\\[0.5cm]
\hline
\end{tabular}
\end{center}
\end{table}


As we have pointed out in the previous section, any Hamiltonian ${\cal H}_g$ defined on one of the abovementioned subalgebras $g$  is completely integrable by construction, since the $N$-th coproduct of the Casimir ${\cal C}_g$ provides the extra integral $\cal I$, which completes the set of $(N-2)$ left integrals $C^{(m)}$ coming from the $h_6$ coalgebra. Note that $\cal I$ is a function of the $h_6$ generators and, as a consequence, is in involution with each of the $C^{(m)}$ integrals. Therefore, we can state that the following $N$D Hamiltonians define completely integrable systems:
\bea
&& {\cal H}_{{\cal D}_+}={\cal H}_{{\cal D}_+}(\kk,A_+,B_+) \nonumber\\
&& {\cal H}_{{\cal D}_-}={\cal H}_{{\cal D}_-}(\kk,A_-,B_-) \nonumber \\
&& {\cal H}_{h_4}={\cal H}_{h_4}(\kk,A_-,A_+,M)\nonumber \\
&& {\cal H}_{\overline{\cal G}_+}={\cal H}_{\overline{\cal G}_+}(B_+,A_-,A_+,M) \label{sub1} \\
&& {\cal H}_{\overline{\cal G}_-}={\cal H}_{\overline{\cal G}_-}(B_-,A_-,A_+,M) \nonumber \\
&& {\cal H}_{\overline{\cal E}}={\cal H}_{\overline{\cal E}}(\mu B_+ + \nu B_-,A_-,A_+,M) \nonumber \\
&& {\cal H}_{gl(2)}={\cal H}_{gl(2)}(\kk,B_-,B_+,M) 
\nonumber
\eea
where the $h_6$ generators are taken through their $N$D symplectic realization (\ref{sympn}) and the Hamiltonian functions are arbitrary functions of the corresponding arguments. 
Moreover, all these systems are superintegrable, since the $h_6$ coalgebra provides $(N-3)$ additional and functionally independent integrals of the motion  given by the `right'  integrals $C_{(m)}$.

The explicit form of the Casimir invariant ${\cal C}_g\equiv {\cal I}$ of each subalgebra $g$ is given in table 2 in terms of the $N$-particle symplectic realization  (\ref{sympn}). It is interesting to remark that the two-photon Casimir (\ref{ac})  can be
expressed in terms of the Casimirs of the four subalgebras
$h_3$, $h_4$, $\overline{\cal G}_+$, $\overline{\cal G}_-$ as
\be
{\cal C}_{h_6}=\frac{1}{{\cal C}_{h_3}}\left( {\cal C}_{\overline{\cal G}_+}
{\cal C}_{\overline{\cal G}_-}-{\cal
C}_{h_4}^2\right).
\label{cl}
\ee

Let us finally comment that all of
these subalgebras are also sub-coalgebras since
the same primitive coproduct (\ref{ab})  holds for all of them as Lie--Poisson  algebras, that is, 
$(g,\Delta)\subset (h_6,\Delta)$. In fact, an alternative approach to the integrability of the systems (\ref{sub1}) would be to consider directly the coalgebra construction for the subalgebra in which ${\cal H}_g$ is defined, thus forgetting about the whole $h_6$ scheme. In that case, 
the integrals of   motion would be given by the `left' and `right' $m$-th coproducts of the Casimir of the subalgebra ${\cal C}_g$, say  ${C}_g^{(m)}$ and ${C}_{g,{(m)}}$ ($m=2,\dots,N$), respectively. 

In this way, by taking into account that  ${C}_g^{(2)}$ and ${C}_{g,{(2)}}$ do not vanish in the subalgebra, we would  obtain  (in principle) a maximum number of $(2N-3)$ independent constants of motion for ${\cal H}_g$, and each set of $N$ functions $\{ {C}_g^{(m)}, {\cal H}_g \}$ or $\{ {C}_{g,{(m)}}, {\cal H}_g \}$ would be in involution.
However,  in the case of ${\cal D}_+$, ${\cal D}_-$ and the  two centrally extended $(1+1)$D Galilean
algebras  $\overline{\cal G}_+$ and
$\overline{\cal G}_-$, the right integrals ${C}_{g,{(m)}}$ turn out to be functionally dependent with respect to the left ones, and in these cases the superintegrability of the associated systems can be only derived by making use of the $h_6$ coalgebra construction.

Some of  these subalgebras have been considered previously from the coalgebra method, so that we refer to the various papers on the subject~\cite{BH07, gal, alfonso}. Nevertheless   it is worthy to point out that in the present $h_6$ framework the integrals of motion coming from $gl(2)$ just provide the ones coming from the spherical symmetry (see table~\ref{Table2}). In terms of the $so(N)$ generators $J_{ij}$, (see section 2)  these read
 $$
 C^{(m)}_{gl(2)}=\sum_{1\le i<j}^m J_{ij}^2\qquad  \qquad
  C_{(m),gl(2)}=\sum_{N-m+1\le i<j}^m J_{ij}^2 .
  $$
 Therefore when a Hamiltonian of the type ${\cal H}_{gl(2)}$ is considered the spherical symmetry and its associated superintegrability is recovered as a particular case of the more general $h_6$-coalgebra setting.

\sect{Generator integrability}

Now, let us choose a given  generator $X$ of $h_6$. If we look for all the generators $X_{j}\,\,\,\,  ( j=1,\dots, l$) commuting with $X$ and we look for all the subalgebras $g_{k}\,\,\,(k=1,\dots, t$) containing $X$ as generator, it becomes  obvious that the Hamiltonian constructed through any function of the type
\be
{\cal H}_{X}={\cal H}_X\left( {\cal C}_{g_{1}},\dots,  {\cal C}_{g_{t}}, X,X_{1},\dots, X_{l}\right),
\label{hx}
\ee
where $ {\cal C}_{g_{k}}$ is the Casimir function of the subalgebra ${g_{k}}$, verifies that
\be
\{{\cal H}_{X},X\}=0.
\ee
Moreover, the $N$-th particle symplectic realization of both $X$ and ${\cal H}_{X}$  will Poisson-commute with the two sets of  integrals $ C^{(m)}$  (\ref{c3m}) and    $C_{(m)}$  (\ref{c3mr}). Therefore, the $N$-th symplectic realization of ${\cal H}_{X}$ is a completely integrable $N$D Hamiltonian system (in fact, superintegrable with a total number of $(2N-4)$ integrals of the motion).

As we shall see in the sequel, by taking into account the information concerning the $h_6$ subalgebras that is contained in tables 1 and 2, this result provides in a straightforward way a bunch of new $N$D integrable systems. We stress that particular integrable systems belonging to the three classes of the generically quasi-integrable Hamiltonians  (\ref{eucla}), (\ref{electro}) and  (\ref{geod}), presented in section 2.1, can be straightforwardly identified.

\subsect{Hamiltonians in involution with $\kk$}

Let us start by considering $X\equiv \kk$. It is immediate to check that the only generator that Poisson-commutes with $\kk$ is $M$, the central one. On the other hand, $\kk$ is contained in the subalgebras ${{\cal D}_{+}},{{\cal D}_{-}},{h_{4}}$ and ${gl(2)}$. Therefore, the most general Hamiltonian with $h_6$-coalgebra symmetry and in involution with $\kk$ is
\be
{\cal H}_{\kk}=
{\cal H}_{\kk}\left({\cal C}_{D_{+}},{\cal C}_{D_{-}},{\cal C}_{h_{4}},{\cal C}_{gl(2)},\kk,M\right)
\ee
where ${\cal H}_{\kk}$ is an arbitrary smooth function  ${\cal H}_{\kk}: \mathbb R^6 \rightarrow \mathbb R$.
Now,   if we take the $N$D symplectic realization of $\mathcal{H}_{\kk}$, we obtain an $N$D integrable Hamiltonian with $(N-1)$ integrals of   motion in involution given by (\ref{c3m}) together with
\be
{\cal I}\equiv \kk = \sum\limits_{i=1}^{N}q_{i}p_{i}.
\ee 

Next, in order to classify the type of Hamiltonian systems that can be constructed  from $\mathcal{H}_{\kk}$, we have to realize that the symplectic realization of $\kk$ and ${\cal C}_{h_{4}}$ are linear in the momenta, while ${\cal C}_{gl(2)}$ is quadratic in $p$. On the other hand, ${\cal C}_{D_{+}}$ is rational in $p$ (for arbitrary $\la_i$) and ${\cal C}_{D_{-}}$ is a rational function in the canonical coordinates $q$. With these ingredients in mind and by considering the three families of systems given in section 2.1, a family of {\em completely integrable geodesic flows} on  $N$D curved spaces is obtained through a choice of the Hamiltonian $\mathcal{H}_{\kk}$ leading to a quadratic homogeneous function in the momenta. Namely, the most general possibility of this type turns out to be
\be
 \mathcal{H}_{\kk}= {\cal C}_{gl(2)}\mathcal{F}({\cal C}_{D_{-}}) +\left(\kk+\frac{M}{2}\right)^{2}
\mathcal{G}\left({\cal C}_{D_{-}}
\right) 
+{\cal C}_{h_{4}}^{2}\rr({\cal C}_{D_{-}}) 
 +\left(\kk+ \frac{M}{2} \right){\cal C}_{h_{4}} \sss(C_{D_{-}}) 
\ee
where ${\cal F}$, ${\cal G}$, $\rr$ and $\sss$ are arbitrary functions.
When this Hamiltonian is written in terms of canonical coordinates we get
\begin{eqnarray}
&&\mathcal{H}_{\kk}= \left(\sum\limits_{1\le i<j}^{N}\,
\left(q_{j}p_{i}-q_{i}p_{j}\right)^{2}\right)
\mathcal{F}\left( 
\dfrac{\left(\sum\limits_{i=1}^{N}\la_{i}q_{i}\right)^{2}}{\sum\limits_{i=1}^{N}q_{i}^{2}}
\right)
+ \left(\sum\limits_{i=1}^{N}q_{i}p_{i}\right)^{2}\mathcal{G}\left(
\dfrac{\left(\sum\limits_{i=1}^{N}\la_{i}q_{i}\right)^{2}}{\sum\limits_{i=1}^{N}q_{i}^{2}}
\right) \notag \\
&&\qquad\qquad + \left(\sum\limits_{1\le i <j}^{N}\left(\la_{j}p_{i}-\la_{i}p_{j}\right)\left(\la_{j}q_{i}-\la_{i}q_{j}\right)\right)^{2}\!\!
 \rr\!\! \left(
\dfrac{\left(\sum\limits_{i=1}^{N}\la_{i}q_{i}\right)^{2}}{\sum\limits_{i=1}^{N}q_{i}^{2}}
\right)
 \notag \\
&& \qquad\qquad
+\left(\sum\limits_{1\le i <j}^{N}\left(\la_{j}p_{i}-\la_{i}p_{j}\right)\left(\la_{j}q_{i}-\la_{i}q_{j}\right)\right)\left(\sum\limits_{i=1}^{N}q_{i}p_{i}\right) \sss\!\! \left(
\dfrac{\left(\sum\limits_{i=1}^{N}\la_{i}q_{i}\right)^{2}}{\sum\limits_{i=1}^{N}q_{i}^{2}}
\right)
\end{eqnarray}
which is an $N$D integrable geodesic flow that depends on four arbitrary functions and $N$ free parameters $\la_i$.


\subsect{Hamiltonians in involution with $A_+$}

When the generator $A_+$ is considered, we find that both $B_{+}$ and $M$ commute with it. On the other hand, $A_+$ belongs to the subalgebras ${{\cal D}_{+}},{h_{4}},{\overline{\cal G}_+},{\overline{\cal G}_-}$ and $\overline{\cal E}$. Both facts lead to the completely integrable Hamiltonian.
\be
\mathcal{H}_{A_{+}}= \mathcal{H}_{A_{+}}\left( {\cal C}_{{\cal D}_{+}},{\cal C}_{h_{4}},{\cal C}_{\overline{\cal G}_+},{\cal C}_{\overline{\cal G}_-},{\cal C}_{\overline{\cal E}},A_{+},B_{+},M \right) 
\ee
whatever the function $\mathcal{H}_{A_{+}}$ is.
In this case, the appearance of $B_+$ (the Euclidean kinetic energy term) allows for a wider set of possibilities. In particular, all the following types of integrable Hamiltonians can be considered as specific cases of $\mathcal{H}_{A_{+}}$ for which the remaining integral is  ${\cal I}\equiv A_+ = \sum\limits_{i=1}^{N}\la_{i}p_{i}$. This integral can be interpreted as a $\la_i$-generalization of the translational symmetry.

\medskip

\noindent $\bullet$ {\em Natural Hamiltonians.} The only possibility is
\be
\mathcal{H}_{A_{+}}=\dfrac12\, {B_{+}}+{\cal F}\left({\cal C}_{\overline{\cal G}_-}\right)
=\sum\limits_{i=1}^{N}\dfrac{p_{i}^{2}}{2}+{\cal F}
\left(\sum\limits_{1\le i<j}^{N}\,\left(\la_{j}q_{i}-\la_{i}q_{j}\right)^{2} 
\right) .
\ee
Note that this Hamiltonian is {\em not} defined within the $\overline{\cal G}_-$ subalgebra. In the $N=2$ case, the Calogero--Moser systems~\cite{Ca71, Mo75} arise as particular choices for ${\cal F}$.

\medskip
\noindent $\bullet$ {\em Electromagnetic Hamiltonians.} We can add linear terms in the momenta to the previous Hamiltonian leading to
\bea
&&\!\!\!\!\!\!\!\!\! \mathcal{H}_{A_{+}}=\frac 12\, {B_{+}}+{\cal C}_{h_{4}}\,{\cal G}\left( {\cal C}_{\overline{\cal G}_-}\right)+A_{+}\rr\left( {\cal C}_{\overline{\cal G}_-}\right)+{\cal F}\left( {\cal C}_{\overline{\cal G}_-}\right) \nonumber\\
&&\!\!\!\!\!\! \!\!\! \qquad \   =\sum\limits_{i=1}^{N}\dfrac{p_{i}^{2}}{2}+\left(
\sum\limits_{1\le i <j}^{N}\left(\la_{j}p_{i}-\la_{i}p_{j}\right)\left(\la_{j}q_{i}-\la_{i}q_{j}\right)
\right){\cal G}\left(\sum\limits_{1\le i<j}^{N}\,\left(\la_{j}q_{i}-\la_{i}q_{j}\right)^{2}
\right) \notag \\
&&\!\!\!\!\!\!\!\!\! \qquad \quad +\left(\sum\limits_{i=1}^{N}\la_{i}p_{i}\right)\,\rr\left(\sum\limits_{1\le i<j}^{N}\,\left(\la_{j}q_{i}-\la_{i}q_{j}\right)^{2}\right)
+{\cal F}\left(\sum\limits_{1\le i<j}^{N}\,\left(\la_{j}q_{i}-\la_{i}q_{j}\right)^{2}\right).
\end{eqnarray}

\noindent $\bullet$ {\em Geodesic flows.} The  most general expression coming from $\mathcal{H}_{A_{+}}$ and with homogeneous quadratic dependence in the momenta is given by
\bea
&&\!\!\!\!\!\!\!\!\!\!\!\!\!\!\!\!\!\!\!\!
{\cal H}_{A_{+}}= {\cal C}_{h_{4}}^{2}{\cal F}\left( {\cal C}_{\overline{\cal G}_-} \right)+{\cal C}_{\overline{\cal G}_+}  {\cal G}\left( {\cal C}_{\overline{\cal G}_-}\right)
+B_{+}  \rr\left(  {\cal C}_{\overline{\cal G}_-} \right)+A_{+}^2{\cal S} \left( {\cal C}_{\overline{\cal G}_-}\right) +A_{+}{\cal C}_{h_{4}}{\cal T} \left( {\cal C}_{\overline{\cal G}_-}\right)\nonumber\\
&&\!\!\!\!\!\!\!=
 \left(\sum\limits_{1\le i <j}^{N}\left(\la_{j}p_{i}-\la_{i}p_{j}\right)\left(\la_{j}q_{i}-\la_{i}q_{j}\right)
\right)^{2}\,{\cal F}\left(\sum\limits_{1\le i<j}^{N}\,\left(\la_{j}q_{i}-\la_{i}q_{j}\right)^{2}\right)
\notag \\
&&\!\!\!\!\!+ \left(
\sum\limits_{1\le i<j}^{N}\,\left(\la_{j}p_{i}-\la_{i}p_{j}\right)^{2}
\right)\,{\cal G}\left(\sum\limits_{1\le i<j}^{N}\,\left(\la_{j}q_{i}-\la_{i}q_{j}\right)^{2}\right) \notag \\
&&\!\!\!\!\!+\left(\sum\limits_{i=1}^{N}p_{i}^{2}\right)\rr\left(
\sum\limits_{1\le i<j}^{N}\,\left(\la_{j}q_{i}-\la_{i}q_{j}\right)^{2}
\right)  
+ \left(\! \sum\limits_{i=1}^{N}\la_{i}p_{i}\!\! \right)^2
 {\cal S}\! \left(\sum\limits_{1\le i<j}^{N}\,\left(\la_{j}q_{i}-\la_{i}q_{j}\right)^{2}\right)
  \notag  \\
&&\!\!\!\!\!+ \left(\! \sum\limits_{i=1}^{N}\la_{i}p_{i}\!\! \right) \left( \sum\limits_{1\le i <j}^{N}\!\! \left(\la_{j}p_{i}-\la_{i}p_{j}\right) \left(\la_{j}q_{i}-\la_{i}q_{j}\right)
\right) {\cal T}\! \left(\sum\limits_{1\le i<j}^{N}\,\left(\la_{j}q_{i}-\la_{i}q_{j}\right)^{2}\right).
\end{eqnarray}


\subsect{Hamiltonians in involution with either $A_-$, $B_-$ or $B_+$}

To end with we present jointly these tree types of Hamiltonians, since all of them provide new examples of $N$D geodesic flows.

\noindent $\bullet$ {\em   $A_-$-Hamiltonians}. If we consider that ${\cal I}\equiv A_- = \sum\limits_{i=1}^{N}\la_{i}q_{i}$, is straightforward to prove that the most general integrable  $A_-$-Hamiltonian reads
\be
{\cal H}_{{A}_{-}}= {\cal H}_{{A}_{-}}\left( {\cal C}_{{\cal D}_{-}},{\cal C}_{h_{4}},{\cal C}_{\overline{\cal G}_{+}},{\cal C}_{\overline{\cal G}_{-}},{\cal C}_{\overline{\cal E}},A_{-},B_{-},M \right).
\ee
In this case, geodesic flow Hamiltonians are available through the particular choice
\be
\mathcal{H_{A_{-}}}={\cal C}_{h_{4}}^{2}\, {\cal F}\left( {\cal C}_{{\cal D}_{-}},   {\cal C}_{\overline{\cal G}_{-}},   A_{-},B_{-}\right)+ {\cal C}_{\overline{\cal G}_{+}}{\cal G}\left({\cal C}_{{\cal D}_{-}},   {\cal C}_{\overline{\cal G}_{-}},   A_{-},B_{-} \right)
\ee
and its $N$-particle symplectic realization can be immediately  obtained.

\noindent $\bullet$ {\em $B_-$-Hamiltonians}. A similar situation is encountered when ${\cal I}\equiv B_-= \sum\limits_{i=1}^{N}q_{i}^2$ is considered. In this case we have that
\be
{\cal H}_{B_{-}}= {\cal H}_{B_{-}}\left( {\cal C}_{{\cal D}_{-}},{\cal C}_{\overline{\cal G}_{-}},{\cal C}_{gl(2)},B_{-},A_{-},M \right)
\ee
and since ${\cal C}_{gl(2)}$ is the only term quadratic in the momenta, we are led to the integrable geodesic flow given by
\be
\mathcal{H}_{B_{-}}={\cal C}_{gl(2)}\, {\cal F} \left({\cal C}_{{\cal D}_{-}},{\cal C}_{\overline{\cal G}_{-}},B_{-},A_{-}  \right).
\ee

\noindent $\bullet$ {\em   $B_+$-Hamiltonians}. Finally, the last possibility is given by ${\cal I}\equiv B_+= \sum\limits_{i=1}^{N}p_{i}^2$.
Now, the most general integrable Hamiltonian is given by
\be
 \mathcal{H}_{B_{+}}=  \mathcal{H}_{B_{+}} \left({\cal C}_{{\cal D}_{+}},{\cal C}_{\overline{\cal G}_{+}},{\cal C}_{gl(2)},B_{+},A_{+},M \right) .
\ee
All the variables fot $\mathcal{H}_{B_{+}}$ (except $M$) depend on the momenta. Therefore, in this case the only integrable geodesic flow can be obtained through 
\be
\mathcal{H}_{B_{+}}=\alpha\,B_{+}+\beta \, A_+^2+\gamma\,{\cal C}_{gl(2)}+\delta\, {\cal C}_{\overline{\cal G}_{+}}
\ee
where $\alpha$, $\beta$, $\gamma$ and $\delta$ are constants.


\sect{Concluding remarks}

As we mentioned in section 3, a third possibility to show the complete integrability for a given  Hamiltonian ${\cal H}$ (\ref{hgen}) with $h_6$-coalgebra symmetry is the direct search for an additional integral ${\cal I}$, that can be assumed to be an unknown function of the $h_6$ generators. An example for this type of construction is given by the following geodesic flow system:
\be
\mathcal{H}=B_{+}\left(\alpha_{1} A_{-}+\alpha_{2} B_{-}+\alpha_{3}\right)
\label{force}
\ee
where $\alpha_1, \alpha_2, \alpha_3$ are non-vanishing constants. Despite its simplicity, this quadratic $h_6$ Hamiltonian neither lives in any $h_6$ subalgebra nor can be included within the cases studied in the previous section. However, the following additional (and functionally independent) constant of   motion can be found by direct computation:
\be
\mathcal{I}=4 \alpha_{1}\alpha_{2}\,A_{+}\left(\kk+\dfrac{M}{2}\right)
+ 4\alpha_{2}\alpha_{3}\,B_{+}
+4 \alpha_{2}^{2}\kk(\kk+M)-\alpha_{1}^{2}\,{\cal C}_{\overline{\cal G}_{+}}.
\ee
This integral provides the complete integrability of the system (\ref{force}) for any dimension. Note that 
in the limit $\alpha_2\to 0$ the Hamiltonian (\ref{force}) belongs to the subalgebra $\overline{\mathcal{G}}_{+}$, and in that case $\mathcal{I}$ is just the Casimir function for such subalgebra, as it should be.
We stress that this direct search for the remaining integral can be indeed very useful, since it can be quite easily computerized. In fact,  the integral $\mathcal{I}$ can be searched among functions with cubic or higher dependence on the momenta (note that all the integrals that we have presented throughout the paper are, at most, quadratic in the momenta). 

Moreover, nothing prevents that, although for a certain ${\cal H}$ defined on $h_6$ the additional integral does exist, such ${\cal I}$ cannot be written as a function (\ref{Iinv}) of the $h_6$ generators ({\em i.e.}, ${\cal I}$ it would not be coalgebra-invariant). This implies that in this case all the previous methods are not applicable and the explicit form for ${\cal I}$ has to be found for each dimension $N$, which constitutes a much more cumbersome task. Nevertheless, we would like to emphasize that any $h_6$-coalgebra invariant Hamiltonian ${\cal H}$ of the form (\ref{hgen}) is only one integral away from being completely integrable, and the search of the complete list of integrable choices for ${\cal H}$ certainly deserves further work.

On the other hand, as it was pointed out from the very beginning of the coalgebra approach to integrability, the existence of the coalgebra symmetry of ${\cal H}$ allows for the direct  construction of integrable deformations of ${\cal H}$ through the use of $q$-deformations of the underlying Poisson coalgebra~\cite{BR}. In the case of $h_6$,
quantum two-photon/Schr\"odinger
algebras have been constructed~\cite{twop,twop2} and its Poisson versions could be used to provide integrable deformations~\cite{quasi} of some of the systems here presented. But here it is important to recall that quantum deformations do not preserve -in general- all the sub-coalgebra structures that exist in the undeformed coalgebra (see~\cite{CP}). So once a given $q$-deformation of $h_6$ had been constructed, the full analysis given in this paper has to be repeated step by step in order to elucidate which integrability properties have survived under $q$-deformation.

Finally, we mention that some of the new families of integrable $N$D Hamiltonians that we have presented in this paper deserve an individual analysis of its dynamical, geometric and physical features, that we also plan to develop in the near future.


\section*{Acknowledgements}

This work was partially supported by the Spanish MICINN  under grant  MTM2007-67389 (with EU-FEDER support), by Junta de Castilla y
Le\'on  (Project GR224) and by INFN-CICyT. A. Blasco acknowledges INFN support during his stay in Roma Tre University, where part of this work was completed.


\end{document}